\documentclass[letterpaper, 10 pt, conference]{ieeeconf}  % Comment this line out
\IEEEoverridecommandlockouts                              % This command is only
\usepackage{graphicx}
\usepackage{amsmath} % assumes amsmath package installed
\usepackage{amssymb}
\usepackage{eurosym} % assumes amsmath package installed
\usepackage{booktabs}
\usepackage[switch]{lineno}
\usepackage{xcolor}
\usepackage{color,soul}
\usepackage{acronym}
\usepackage[noadjust]{cite} %citazioni compatte

\usepackage[absolute]{textpos}  

\newcommand{\bdm}{\begin{displaymath}}
\newcommand{\edm}{\end{displaymath}}
\newcommand{\bi}{\begin{itemize}}
\newcommand{\ei}{\end{itemize}}

\newcommand{\bc}{\begin{center}}
\newcommand{\ec}{\end{center}}
\newcommand{\be}{\begin{equation}}
\newcommand{\ee}{\end{equation}}
\newcommand{\bma}{\begin{math}}
\newcommand{\ema}{\end{math}}
\newcommand{\bea}{\begin{eqnarray}}
\newcommand{\eea}{\end{eqnarray}}
\newcommand{\ba}{\begin{align}}
\newcommand{\ea}{\end{align}}
\newcommand{\bal}{\begin{aligned}}
\newcommand{\eal}{\end{aligned}}
\newcommand{\barr}{\begin{array}}
\newcommand{\earr}{\end{array}}

\DeclareMathOperator*{\argmin}{arg\,min}

\title{\LARGE \bf
On the Control of Energy Storage Systems for Electric Vehicles \\ 
Fast Charging in Service Areas}

\author{Alessandro Di Giorgio, Francesco Liberati, Roberto German\`a, Marco Presciuttini, \\ Lorenzo Ricciardi Celsi, Francesco Delli Priscoli
\thanks{This work is partially supported by the Sapienza-Ateneo 2013 project "Planning and control of flexible electricity demand and generation from renewable energy sources in Smart Grids", no. C26A13LYTB, and the Sapienza-Ateneo 2014 project "ESSSGEM - Energy Storage Systems (ESS) for Smart Grids (SG) Energy Management (EM). ESS control strategies for the optimal energy management of a SG including dispersed renewable energy generators and non-linear loads, as plug-in Electric Vehicles", no. C26A14YT4X.}% <-this % stops a space
\thanks{
A. Di Giorgio, R. German\`a, M. Presciuttini, L. Ricciardi Celsi and F. Delli Priscoli are with the Department of Computer, Control and  Management Engineering, at "Sapienza" University of Rome, Via Ariosto 25, 00185, Rome, Italy, e-mail {\tt\small digiorgio@dis.uniroma1.it}}
\thanks{
F. Liberati is with the SMART Engineering Solutions \& Technologies (SMARTEST) Research Center, eCampus University, Via Isimbardi 10, 22060, Novedrate CO, Italy}
}

\begin{document}

\maketitle
\thispagestyle{empty}
\pagestyle{empty}

%%%%%%%%%%%%%%%%%%%%%%%%%%%%%%%%%%%%%%%%%%%%%%%%%%%%%%%%%%%%%%%%%%%%%%
% ACRONYMS 
\acrodef{DER}{Distributed Energy Resources}
\acrodef{DSO}{Distribution System Operator}
\acrodef{ESS}{Energy Storage System}
\acrodef{MPC}{Model Predictive Control }
\acrodef{NLP}{Non Linear Programming}
\acrodef{PLC}{Power Line Communications}
\acrodef{QCP}{Quadratic Constrained Programming}
\acrodef{QP}{Quadratic Programming}
\acrodef{RES}{Renewable Energy Sources}

%%%%%%%%%%%%%%%%%%%%%%%%%%%%%%%%%%%%%%%%%%%%%%%%%%%%%%%%%%%%%%%%%%%%%%
% ABSTRACT	& KEYWORDS
\begin{abstract}
%This paper presents MAX 15 RIGHE 
This paper presents a real time control strategy for energy storage systems integration in electric vehicles fast charging applications combined with generation from intermittent renewable energy sources. A two steps approach taking advantage of the model predictive control  methodology is designed on purpose to optimally allocate the reference charging power while managing the priority among the plugged vehicles and then control the storage for efficiently sustaining the charging process. Two different use cases are considered: in the former the charging area is disconnected from the grid, so that the objective is to minimize the deviation of electric vehicles charging power from the nominal value; in the latter the focus is on the point of connection to the grid and the need of mitigating the related power flow. In both cases the fundamental requirement for feasible control system operation is to guarantee stability of the storage's state of charge over the time. Simulation results are provided and discussed in detail, showing the effectiveness of the proposed approach.
\end{abstract}

%\begin{IEEEkeywords}
%Electric Vehicles; Fast Charging; Energy Storage Systems, Model Predictive Control; Smart~Grid.
%\end{IEEEkeywords}

\IEEEpeerreviewmaketitle

%%%%%%%%%%%%%%%%%%%%%%%%%%%%%%%%%%%%%%%%%%%%%%%%%%%%%%%%%%%%%%%%%%%%%%
% NOMENCLATURE

\section*{Nomenclature} \label{Nomenclature}
\begin{table}[h]
\small
\begin{tabular}{p{0.5cm} p{7,5cm}}
	$M_t$   			& Set of vehicles to be controlled at time $t$ \\
  $P^{g}$   			& Active power flowing from the grid to the service area \\
  $P^{cs}$   			& Power delivered to the PEVs by charging station\\
	$\hat{P}^{cs}$   			& Maximum power deliverable by charging station\\
  $\bar{\textbf{P}}$   & Vector of PEVs reference charging power levels\\
  $P^{pv}$       		& Power generated by the photovoltaic panel \\
  $P^s$           & Power delivered by the storage\\
  $\hat{P}^s$			& Maximum power deliverable by storage\\
	$\check{\textbf{P}}$	& Vector of PEVs minimum allowed charging levels \\
  $\textbf{P}$	& Vector of PEVs decided charging levels \\
  $\hat{\textbf{P}}$	& Vector of PEVs maximum allowed charging levels \\
  $T$					& Sampling time \\
	$t^{arr}_m$					& Arrival time of the m\textit{th} vehicle \\
  $y$					& ESS state of charge \\
  $\hat{y}$			& Maximum allowed ESS state of charge\\
	$x_m$   			& State of charge of the m\textit{th} vehicle \\
	$\hat{x}_m$   			& Maximum allowed state of charge of the m\textit{th} vehicle \\
\end{tabular}
\end{table}

%% Copyrigth note IEEE
\begin{textblock}{0.8}[0.5,0.5](0.5,0.96) 
\begin{center} 
\noindent\small{\emph{Accepted for publication to the proceedings of the 24th Mediterranean Conference on Control and Automation (MED16).
© 2016 IEEE. Personal use of this material is permitted. Permission from IEEE must be obtained for all other uses, in any current or future media, including reprinting/republishing this material for advertising or promotional purposes, creating new collective works, for resale or redistribution to servers or lists, or reuse of any copyrighted component of this work in other works.}} 
\end{center} 
\end{textblock}

%%%%%%%%%%%%%%%%%%%%%%%%%%%%%%%%%%%%%%%%%%%%%%%%%%%%%%%%%%%%%%%%%%%%%%
% 1. INTRODUCTION
\section{Introduction}\label{sez:intro}
%\hl{NOTA: reference per il paper EEEIC }\cite{ADG_EEEIC15}
%\IEEEPARstart{U}{nlike} other types \cite{MACHOWSKI_BOOK11}.
%1 PAGINA MAX. In this paper, a \ac{MPC} strategy is presented 
As electric vehicle (EV) technology is expected to increase its share in mobility market, technical improvements of fast charging stations (FCSs) and the related geographical coverage represent key factors to allow green travels over large distances. As a matter of fact the Connecting Europe Facilities (CEF) initiative \cite{paper1} co-financed by the European Union TEN-T (Trans European Transport Network) programme, major EV manufacturers and leading energy companies is making investments to support four projects to build a network involving 429 multi-standard fast charging stations (FCSs) across 10 European countries. 
Each project covers a specific area: Rapid Charge Network (UK and Ireland), Central European Green Corridor (Austria, Slovakia, Slovenia, Bavaria and Zagreb), Corri-Door (France), Greening-NEAR (Denmark and Sweden). Full accomplishment will facilitate longer-distance zero-emissions travel covering over 10540 km of Europe's major highways, encouraging use of EVs even in extra-urban contexts. Each FCS is able to charge EVs equipped with the most common technologies such as: CCS (Combined Charging System), CHAdeMO and AC systems.

Despite the expected growth of EVs will produce a strong reduction of $CO_2$ emissions \cite{paper2}, the fast charging concept raises several challenges ranging from the impact of EVs fast charging on the distribution network to business models, as analyzed in \cite{paper3}. On the one hand the installation of a FCS requires to properly size the point of connection (POC) to the grid and face the related start-up and operational costs which have to be evaluated against the expected number of charging sessions over the expected FCS life time. On the other EV charging implies a considerable pulsating load, already investigated and recognized in the relevant literature as detrimental factor for reliable power system operations \cite{paper4} \cite{paper5}; furthermore, simultaneous fast charge sessions lead to a huge increment of peak power demand, stressing strongly the capacity of the grid, benefits of energy storage systems (ESSs) are analyzed in order to mitigate these effects \cite{paper6}. This is the reason why a reasonable scenario for sustainable large scale deployment of FCSs consider their installation in combination with renewable energy sources (RES) and controlled ESSs for locally matching demand and supply, then allowing to downsize the nominal power at POC and mitigate the impact on the medium voltage distribution grid. Taking into account this reference plant, in this paper a real time strategy is proposed to control the ESS in two different cases, as enabler of charging at minimum time and minimum impact on the grid respectively.  
The remainder of the paper is organized as follows. Section \ref{sez:scenario} presents the reference scenario, in which the status of service area connection to the grid plays the role of discriminant between stand alone and grid connected case. Section \ref{sez:problem} details the formalization of the problem in the most general case, covering both addressed use cases. Section \ref{sez:results} shows and discusses the performed simulations, while conclusions and future works are presented in Section \ref{sez:conclusions}. 
%%%%%%%%%%%%%%%%%%%%%%%%%%%%%%%%%%%%%%%%%%%%%%%%%%%%%%%%%%%%%%%%%%%%%%
% 2. STATE OF THE ART
\section{State of the art } \label{sez:soa}
%The above-mentioned problems are strictly related to the problem of critical infrastructure control, extensively dealt with in the literature ranging from theoretical studies [A], [B] to practical solutions [C], [D]. More specifically, optimal location and size of FCSs contribute to cost reductions....

The problems above mentioned are subject of studies in literature. Optimal location and size of FCSs contribute to cost reductions. In \cite{paper7}, characterized by a MINLP optimization approach, results are sensible to the correct inclusion of EV and electric grid losses in the problem, while in \cite{paper8}, the total cost of installation of fast charging station and loss on distribution are proposed as objective functions of the optimization model, and robustness and effectiveness characterize the proposed algorithm. 
The importance of ESS in peak power demand problem, is shown by authors in \cite{paper9}, in which they have introduced a policy involving the future power demand by the expected EVs: this method does not achieve the optimal costumers' satisfaction, but it works in a "best effort" way, guaranteeing fairness during insufficient supply energy.
In \cite{paper10} authors show how ESS contributes also in economic field: through a MINLP formulation which takes into account the capital costs of the transformer, distribution feeder and ESS for the FCS, authors extract the optimal size of ESS, achieving a total costs reduction over $20 \%$. 
In \cite{paper11} authors propose a control strategy, taking into account the historical charging load data: considering the average charging load and the time-of-use electricity price, optimal charging/discharging period for ESS is used to reduce costs about $10 \%$. 
The relevance of ESS covers several aspects: in \cite{paper12} the model provided by authors is centered on EVs charging schedule based on day-ahead considerations; ESS is used to mitigate uncertainty and inaccurate prediction.
In \cite{paper13} the proposed control strategy (based on the distributed bus signaling method) utilizes a dedicated ESS which compensates some negative effects of fast charging session, providing ancillary services to the grid without
affecting predefined charging algorithm of EV battery, extending in this way their life-time. In \cite{paper14} ESS is used as an energy buffer between the grid and the electric vehicle to recharge; the control strategy is characterized by a hierarchical structure composed by an active and reactive power control, a state of charge ESS balancing control and an EVs charge control. Results highlight how just the $30 \%$ of power involved in charging sessions is provided by the grid.
In this paper a real time control strategy is presented for energy storage system integration in electric vehicles fast charging applications combined with generation from photovoltaic panels. The characterizing aspect of the work is a formulation of the control problem taking into account simultaneous requirements from the drivers, asking for a minimum time charging service, and the distribution system operator, interested in mitigating the impact of EVs charging on the grid while increasing RES hosting capacity.  

%MAX MEZZA COLONNA CON IDICAZIONE DEGLI ELEMENTI INNOVAIVI DEL LAVORO. DA INTEGRARE SUCCESSIVAMENTE NELL'INTRODUZIONE.
%%%%%%%%%%%%%%%%%%%%%%%%%%%%%%%%%%%%%%%%%%%%%%%%%%%%%%%%%%%%%%%%%%%%%%
% 3. REFERENCE SCENARIO
\section{Reference scenario} \label{sez:scenario}
\begin{figure}[t!]
\centering
\includegraphics[clip,width=0.95\columnwidth]{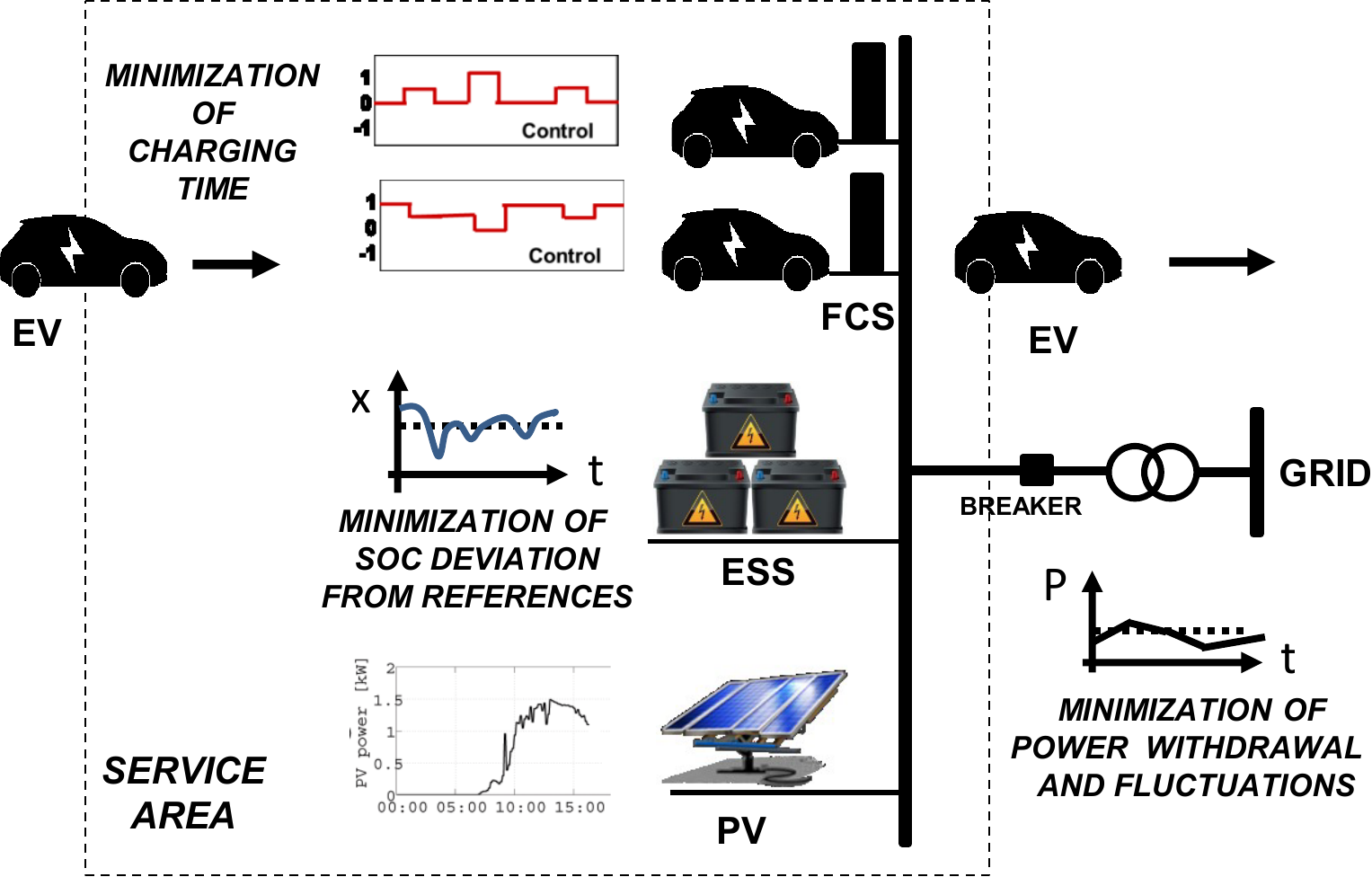}
\caption{Reference scenario.}
\label{fig:scenario}
\end{figure}
%
%\bigskip
%\begin{figure}[t]
%\centering
%\includegraphics[width=\columnwidth]{imm/Scenario.pdf}
%\caption{Reference scenario.}
%\label{fig:scenario}
%\end{figure}
%1 COLONNA MASSIMO COMPRENSIVA DI IMMAGINE. 
The reference scenario at the basis of this study is constituted by a service area in which a fast charging station, a photovoltaic panel and an ESS are included, as shown in fig.\ref{fig:scenario}. The FCS allow to charge the EVs at different power levels, including the typical 22 kW public level and higher. The power which comes from the PV panel is stored by the ESS, which in turn release power during the execution of the charging sessions.   
Grid structure is reduced to the POC and is modeled as an infinite bus, such that voltage and frequency can be considered equal to their nominal values. Two cases are presented using the above scenario: 
\bi
\item in the \emph{grid connected} case, the EVs are recharged at nominal power, which is delivered both by the ESS and the grid, so that a fundamental requirement is to find the optimal trade-off between the need of avoiding full ESS charging or discharging and the one of mitigating the power flow at POC level; 
\item  in the \emph{stand alone} case the service area is electrically disconnected from the grid, so that EVs are recharged by the power stored in the ESS, with the possibility of dinamically modulating the charging power over time, then leading to the need of avoiding full ESS charging or discharging while minimizing the deviation of EVs charging power from the nominal value.
\ei
%%%%%%%%%%%%%%%%%%%%%%%%%%%%%%%%%%%%%%%%%%%%%%%%%%%%%%%%%%%%%%%%%%%%%%
% 3. Control Problem formalization
\section{Control problem formalization} \label{sez:problem}
Since the two above mentioned cases are based on the the same plant, sharing the presence of PV and ESS and differentiating by the supplementary source represented by the grid, the most general problem will be formalized, and then detailed for each case.
A first step in the problem formalization requires the definition of proper charging setpoints for the EVs. As a matter of fact, in some situations the power required by the EVs could exceed the power that can be supplied by the charging station, denoted with $\hat{P}^{cs}$. In fact, while in the grid connected case the required power will be fully provided by the storage and the grid, in the stand alone case the power delivered to the EVs will depend on the current state of charge of the storage (which could be not enough to serve all the ongoing charging sessions). In both cases, the proposed EV control scheme foresees the repetition of both the charging setpoint definition step and the MPC control step. In fact, both have to be executed in order to make the control approach effective against variations of the renewable output, new vehicles arrival at the charging station, mismatches between the charging levels commanded by the charging station and the actual ones, etc. 
The next subsection details the problem of charging setpoints definition.
\subsection{Charging Setpoints Definition Problem} 
\label{sez:subproblem1}
Let us denote with $M_t$ the set of vehicles to be controlled at time $t$ and with $x_m(t)$ and $\hat{x}_{m}$, respectively, the current state of charge and the maximum state of charge of the generic vehicle $m\in M_t$. 

For each vehicle, let us introduce the charging rate
\begin{equation}
\bar{P}_m(t)=\frac{\hat{x}_{m}-x_{m}(t)}{T}, \ \ \forall m\in M_t
\end{equation} 
where $T$ is the sampling time. If $\bar{P}_m(t)$ is not contained within the allowed maximum and minimum charging limits $\hat{P}_m$ and $\check{P}_m$, it takes the values of the limits themselves:
\begin{equation}\label{Preq1}
\bar{P}_m(t)\leftarrow \begin{cases}
\ \check{P}_m \quad \text{if} \quad \bar{P}_m(t) < \check{P}_m \\
\ \hat{P}_m \quad \text{if} \quad \bar{P}_m(t) > \hat{P}, \ \ \forall m\in M_t \\
\ \bar{P}_m(t) \quad \quad \text{otherwise}
\end{cases}
\end{equation} 
Since the available charging power $\hat{P}^{cs}$ at the charging station could be not enough to recharge all the vehicles at power levels $\bar{P}_m(t)$, a charging setpoint definition step is necessary, as detailed in the following.

A least square problem is defined with the aim of finding a feasible repartition of the available charging power among the charging sessions, in such a way that: (i) the cumulative charging power delivered by the charging station does not exceed $\hat{P}^{cs}$ (feasible maximum power charging); (ii) long waiting times are penalized (thus striving the system towards a balancing of the waiting times). The problem is formalized as follows
\begin{equation}
\bar{\textbf{P}}(t)\leftarrow \argmin_{\textbf{P}}\frac{1}{2}\left[\textbf{P}-\bar{\textbf{P}}(t)\right]^T\textbf{A}\left[\textbf{P}-\bar{\textbf{P}}(t)\right]
\end{equation}
\begin{equation*}
\text{subject to} \quad \textbf{1}^T\textbf{P}\le \hat{P}^{cs}
\end{equation*}
where 
\begin{equation}
\textbf{A}=\left( \begin{matrix} (t-t_1^{arr})^e & \quad & 0 \\ \quad & \ddots & \quad \\ 0 & \quad & (t-t_{|M_t|}^{arr})^e
\end{matrix} \right).
\end{equation} 
$\textbf{1}$ is a vector of ones, $\textbf{P}\in\mathbb{R}^{|M_t|}$ and $\bar{\textbf{P}}(t)\in\mathbb{R}^{|M_t|}$ is the vector whose general m-th component is $\bar{P}_m(t)$. The arrival time of the m-th vehicle is denoted with $t_m^{arr}$; $e$ is a factor to weight the recharging time balancing requirement.

For the sake of simplicity, $\textbf{P}$ is modelled as a vector of continuous variables, while it is known that current reference charging protocols supporting charging station to EV communication only allow for semi-continuous charging variables (i.e., the charging level is either zero, or greater than a \textit{positive} quantity). However, the problem can be easily extended to the semi-continuous case following the approach in \cite{paperIEC} (which is not done in this paper in order not to burden the formulation). 
\subsection{\ac{MPC} Problem} \label{sez:subproblem2}
%Presentazione del problema di tracking in forma complessiva per i casi stand alone e grid connected. Note sulla particolarizzazione del problema generale ai due casi d'uso.%
Once a feasible set of charging setpoints for the ongoing charging sessions is defined, the general \ac{MPC} problem can be formalized as follows
\begin{equation}\label{eq:MPC}
\begin{aligned}
\min\quad &\alpha[y(t+1)-y^{ref}]^2\\
& +\beta\left[\textbf{P}-\bar{\textbf{P}}(t)\right]^TA\left[\textbf{P}-\bar{\textbf{P}}(t)\right]\\
& +\gamma P^{g}(t)^2+\delta\left[P^{g}(t)-P^{g}(t-1)\right]^2 \\
\end{aligned}
\end{equation}
Subject to:
\begin{eqnarray}
&& y(t+1)=y(t)-(1+\epsilon_s)TP^{s}(t)+ TP^{pv}(t) \qquad \\
&& 0 \leq y(t+1)\leq \hat{y}^s \\
&& P^{s}(t)(1+\epsilon_s)\leq \hat{P}^s\\
&& P^{g}(t)+P^{s}(t)=\textbf{1}^T\textbf{P}(t) \\  %piccola imprecisione qui
%&& P^{stor}(t)-\textbf{1}^T\textbf{P}=0\\
&& \check{\textbf{P}}\le\textbf{P}\le \bar{\textbf{P}}(t) %\\
%&& f(x_v(t))=P_{req}^*(t)\\
%&& Binary: \quad S
\end{eqnarray}
in which $\alpha$, $\beta$, $\gamma$, $\delta$ are wight parameters, $y^{ref}$ represents the reference state of charge value set for the storage and $y(t)$ its current state of charge; $P^{pv}$ is the power supplied by the photovoltaic panel (PV); $\epsilon_s$ represents the loss related to the storage, while $P^{s}$ and $\hat{P}^s$ the storage delivered power and the maximum deliverable power, respectively.

The above \ac{MPC} iteration subsume the goals and constraints present in both the standalone case and in the grid connected case. In particular, the objective function is such that the following objectives are balanced: (i) limited excursion of the storage state of charge from its reference value, (ii) charging levels close to the setpoint values defined in the setpoint definition problem, (iii) flattening of the peaks in the power withdrawal from the grid and, finally, (iv) limited variations in the power withdrawal from the grid. Constraints include the dynamics of the storage and the technical limits imposed by the storage itself and by the grid. In particular, condition (6) expresses the balance equation of the storage state of charge; its value is bounded as expressed by (7). The power delivered by the storage can not overcome its maximum value (8) and the CS output power is bounded by a minimum absorption power and the relative reference (10). Condition (9) expresses the power balance condition for the service area.

The above MPC iteration can be specialized as follows in the two case studies dealt with in this paper:
\begin{itemize}
\item \textit{Grid Connected Case}. In the grid connected case, $\beta=0$ in the objective function, since $\textbf{P}=\bar{\textbf{P}}$. In fact, the charging sessions can always be served at maximum power and the goal of the control problem is that of mitigating peaks in the power grid withdrawals by acting on the storage (and avoiding at the same time large excursions of the storage state of charge from its reference value).
\item \textit{Standalone Case}. In the stand alone case, $P^{g}$ is equal to zero. Therefore, the objective function to minimize involves only the first two terms.
\end{itemize}  
%\subsection{Solving procedure} \label{sez:procedure}
The first references generation problem requires the resolution of a nonlinear quadratic system characterized by real variables and linear constraints. Instead, a mixed integer quadratic programming  problem has to be solved for the tracking problem. IBM-Cplex is the solver chosen to address them. Resolution is based on an iterative algorithm: at each step, input information are used to compute the state at the following step.
%INSERIRE LA DESCRIZIONE DEL TIPO DI PROBLEMA E SOLVER UTILIZZATO. 
%%%%%%%%%%%%%%%%%%%%%%%%%%%%%%%%%%%%%%%%%%%%%%%%%%%%%%%%%%%%%%%%%%%%%%
% 5. RESULTS
\section{Results} \label{sez:results}
\subsection{Simulation setup} \label{sez:sim_setup}
%Descrizione del case study. Dimensionamento dei conponenti.
In order to perform simulations, a charging station which is able to deliver 120 kW ($\hat{P}^{cs}$) and charge simultaneously 4 EVs (two 50 kW, one 43 kW and one 22 kW) is considered. A super capacitor storage system is taken into account, with maximum power flow $\hat{P}^s=150 kW$, battery capacity $\hat{y}=300 kWh$ and losses amounted to $\epsilon_s=10\%$; the reference state of charge is set to 50\% of the nominal battery capacity. The PV profile is modeled as a measurable disturbance; its nominal power output is 120 kW, with conversion losses $\epsilon_{pv}= 15\%$. A time window of 120 minutes is analyzed, with sampling time $T=60s$. The weight parameters $\alpha$, $\beta$ and $\gamma$ are put, respectively, equal to 10, $ 5 \times 10^{6}$, $3\times 10^{7}$, while $P_{g}$ ramping rate effect is analyzed in case of $\delta=10$ and $\delta=5 \times 10^{6}$. The time dependency is given setting $e=3$.  

Simulations have been performed using a Notebook Samsung, Intel Core i7, 2.20 GHz,  6 GB RAM, running Windows 8 Pro. The control framework has been built in Matlab R2013a 32 bit, and the optimization problem at the basis of the MPC has been solved at each iteration by using the commercial solver Cplex v.12.6.1.
%\begin{figure}[p!]
%	\centering
%\includegraphics[clip,width=0.98\columnwidth]{Grafici/riferimentoNOTempo.jpg}
%	\caption{Reference generation: time independent case.}
%	\label{fig:ref_ind}
%\end{figure}
%\begin{figure}[h]
%	\centering
%\includegraphics[clip,width=0.98\columnwidth]{Grafici/riferimentoTempo.jpg}
%	\caption{Reference generation: time dependent case.}
%	\label{fig:ref_dip}
%\end{figure}
%
\subsection{Reference generation}
\label{sez:references}
\begin{figure}[t!]
	\centering
	\includegraphics[clip,width=0.99\columnwidth]{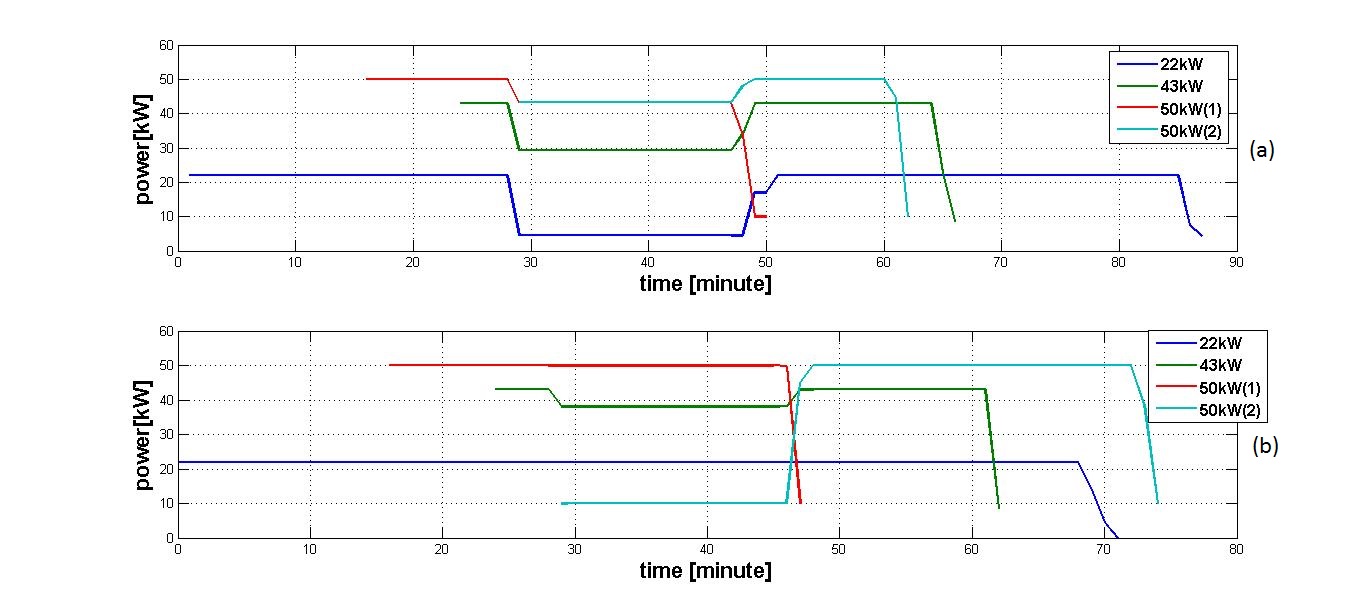}
\vspace{-0.6cm}
	\caption{Reference generation: \textit{(a)} time independent case; \textit{(b)} time dependent case.}
	\label{fig:reference}
\end{figure}
The first simulation is dedicated to the generation of power references for charging, assessing the effect of the temporal component, taking the grid connected case as example; as a matter of fact fig. \ref{fig:reference} reports the references, as resulting from the last iteration of the simulated control strategy.

When $e=0$ (time independent case), the system works giving priority for charging to the EVs connected to the plugs releasing the higher power level, as shown in the subfigure (a).
If the sum of the required power is less than the maximum power deliverable by the system in that moment, each EV is associated with a power reference equal to the nominal charging power. When this condition is violated, a redefinition of the references occurs, as depicted for example at the minute 28.
All EVs are simultaneously connected, therefore the sum of the maximum charging powers exceeds $\hat{P}^{cs}$; the  22 kW charging session is subject to more performance degradation, up to $80 \%$,  with the consequent extension of the expected charging time, exceeding 80 minutes. Assuming the actual charging powers exactly track the calculated references, the EVs with greater power demand are expected to charge within a reasonable time: despite a degradation of $32 \%$, the EV connected to the 43 kW plug leaves the service area in about 40 minutes, while the last two EVs (50 kW) complete the charging operation in about 30 minutes, due to the less performance degradation~($13 \%$). 

When $e\neq 0$ (time dependent case), the performances are subject to the order of arrival;
the higher $e$, the higher the effect. Setting $e=3$ it is already possible to appreciate how the temporal component affects the reference generation, as shown in the subfigure (b). Obviously, by choosing a smaller value the behaviour of the system will be similar to the previous case. The system reserves the best treatment to the first EV arrived (22 kW) and its reference is maintained constant for all the charging time. 
When the system is not able to deliver the nominal required power to each vehicle (minute 45), the latest arrived one (50 kW) receives the worst performance; it is subject to a considerable reduction of the power reference ($80 \%$), which comes back to its nominal value as soon as the system is able to support it (an EV earlier arrived is disconnected at the minute 47).
%Discussione modulare sia del caso con pesi indipendendenti dal tempo che dipendenti dal tempo. In caso di problemi di numero massimo di pagine daremo visibilità al solo caso più significativo.
%
%
%
\subsection{Test case 1: Stand alone plant} \label{sez:testcase1}
\begin{figure}[t]
	\centering
	\includegraphics[clip,width=0.99\columnwidth]{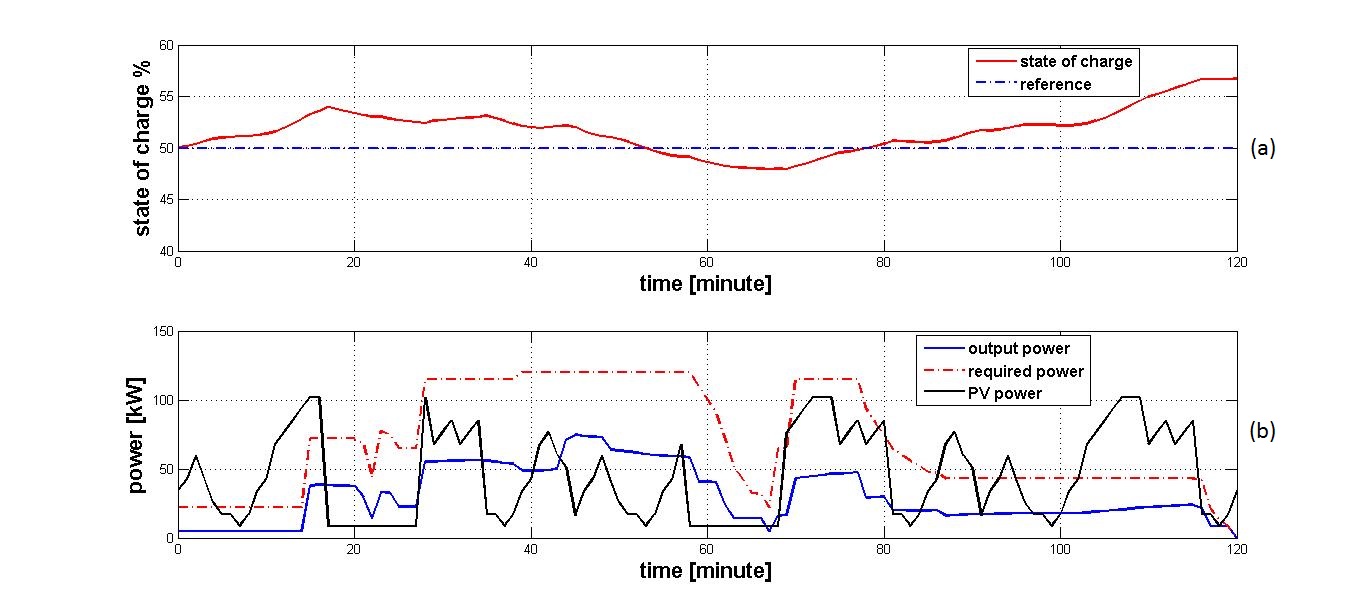}
	\vspace{-0.6cm}
	\caption{Stand alone system in time independent case: \textit{(a)} ESS state of charge and reference; \textit{(b)} output, required and PV power.}
	\label{fig:gen_ind}
\end{figure}
\begin{figure}[t]
	\centering
	\includegraphics[clip,width=0.99\columnwidth]{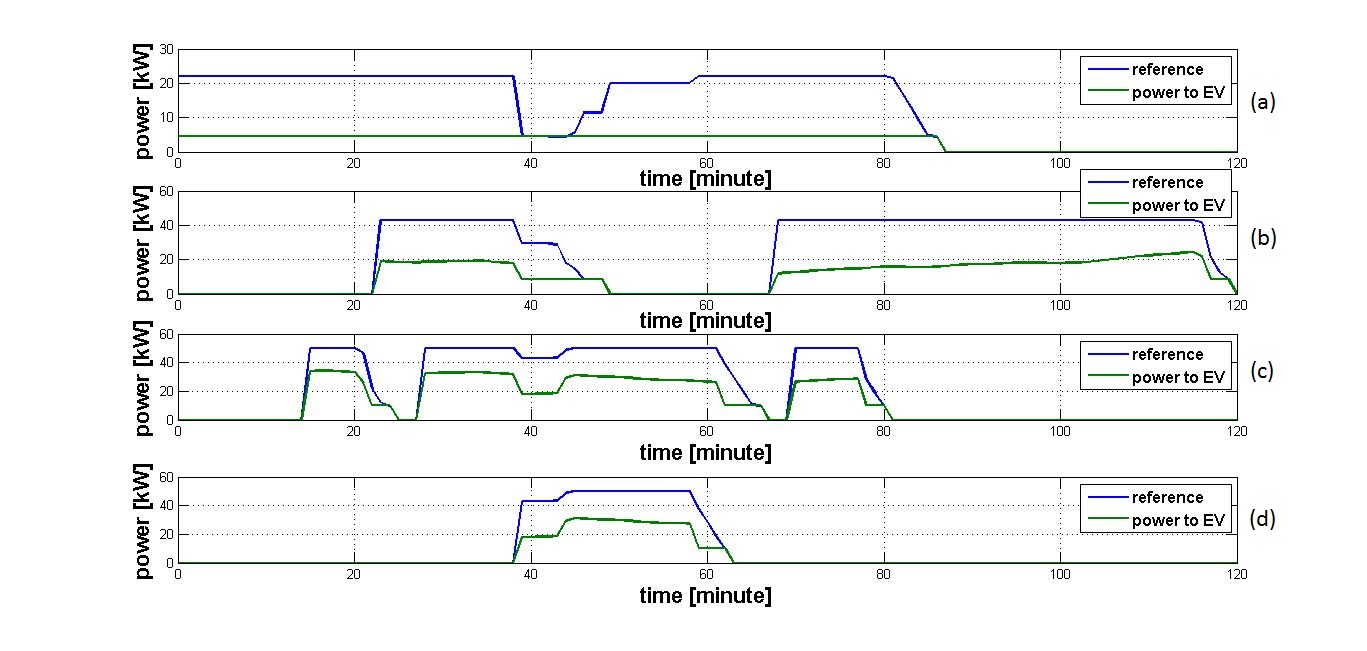}
	\vspace{-0.8cm}
	\caption{Stand alone system in time independent case: reference and power for each plug.}
	\label{fig:sin_ind}
\end{figure}
\begin{figure}[t]
	\centering
	\includegraphics[clip,width=0.99\columnwidth]{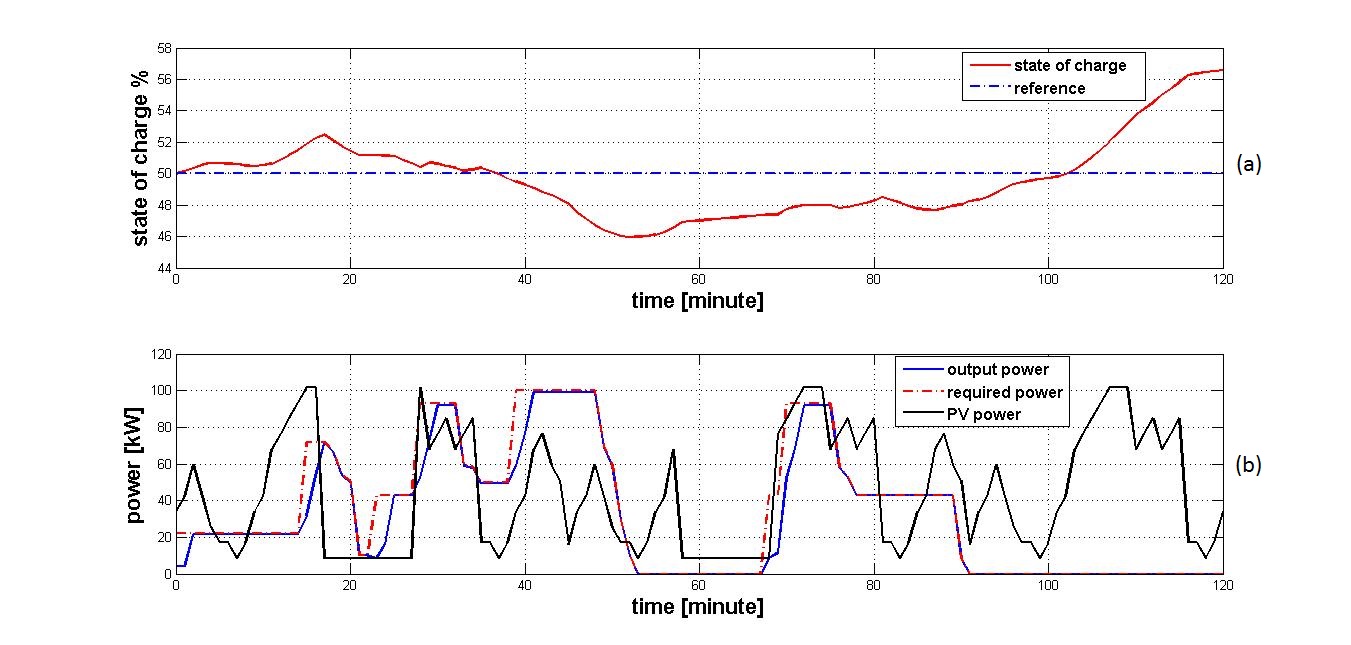}
	\vspace{-0.8cm}
	\caption{Stand alone system in time dependent case: \textit{(a)} ESS's state of charge and reference; \textit{(b)} output, required and PV power.}
	\label{fig:gen_dip}
\end{figure}
\begin{figure}[t]
	\centering
	\includegraphics[clip,width=0.99\columnwidth]{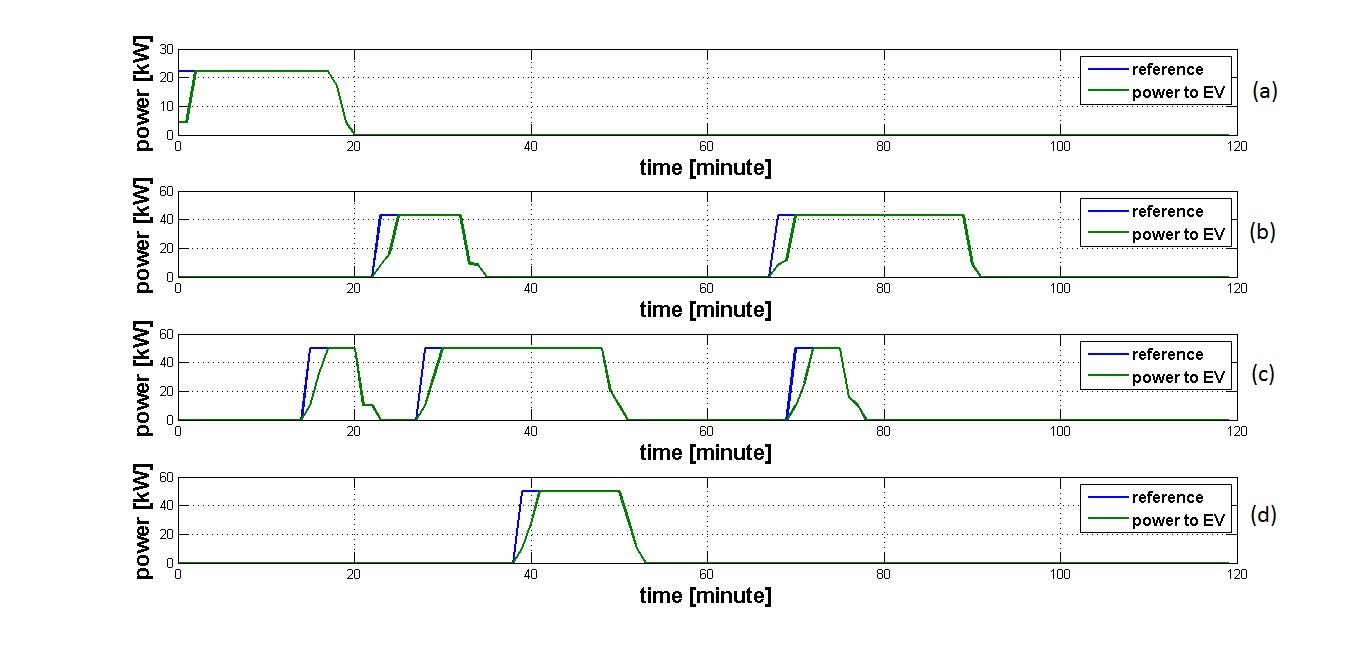}
	\vspace{-0.8cm}
	\caption{Stand alone system in time dependent case: reference and power for each plug.}
	\label{fig:sin_dip}
\end{figure}

Several simulations have been performed varying the initial ESS's state of charge. In the proposed set of simulations, this value is equal to the state of charge reference value. In the time independent case (fig.\ref{fig:gen_ind}) the aggregated profile of the powers delivered by the FCS tracks the shape of the aggregated reference profile, but not the exact value. This behaviour is due to the fact that, with the chosen values of the weights appearing in the target function, maintaining the ESS's state of charge near to its reference has a greater priority than EVs charging. In fig.\ref{fig:sin_ind} the power delivered through each plug is depicted. In the time dependent case (fig.\ref{fig:gen_dip}), the temporal component has also the effect of shifting the weight on the EVs charging, as will be illustrated in section \ref{sez:caso_particolare}: this implies that the EVs receive a better treatment and the nominal charging power is fully delivered to each vehicle, as depicted in fig.\ref{fig:sin_dip}. As a consequence, it is interesting to note that not only the charging operations end 30 minutes in advance with respect to the time independent case, but also the critical situation in which the aggregated required power is greater than the maximum deliverable by the FCS is avoided, since each EV leaves the service area in a short time and therefore it does not create a waiting queue; this result is obtained at the cost of allowing a larger excursion of the ESS's state of charge with respect to its reference in the time independent case, which however remains within the acceptable operating range.
\subsection{Test case 2: Grid connected plant} \label{sez:testcase2}
\begin{figure}[t]
	\centering
	\includegraphics[clip,width=0.99\columnwidth]{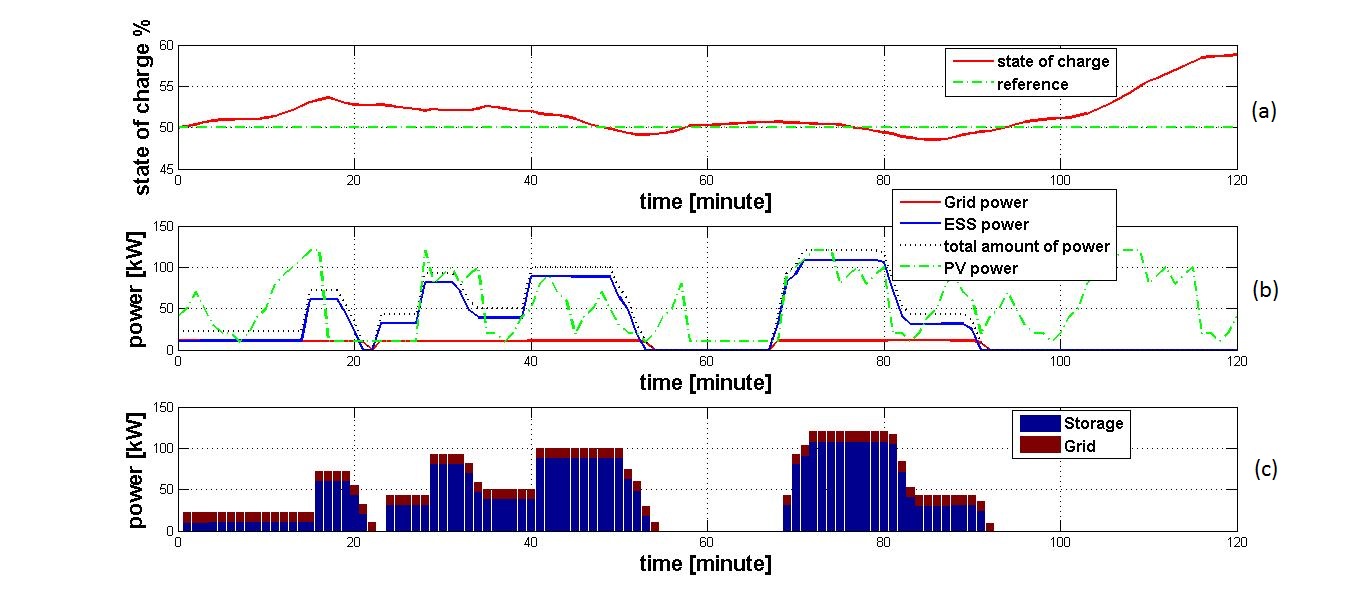}
	\vspace{-0.8cm}
	\caption{Grid connected system in time dependent case with medium level of storage charge and low level of $\delta$: \textit{(a)} ESS's state of charge and reference, \textit{(b)} Grid, ESS and PV power and total power, \textit{(c)} ESS and grid power}
	\label{fig:low_delta}
\end{figure}
\begin{figure}[t]
	\centering
	\includegraphics[clip,width=0.99\columnwidth]{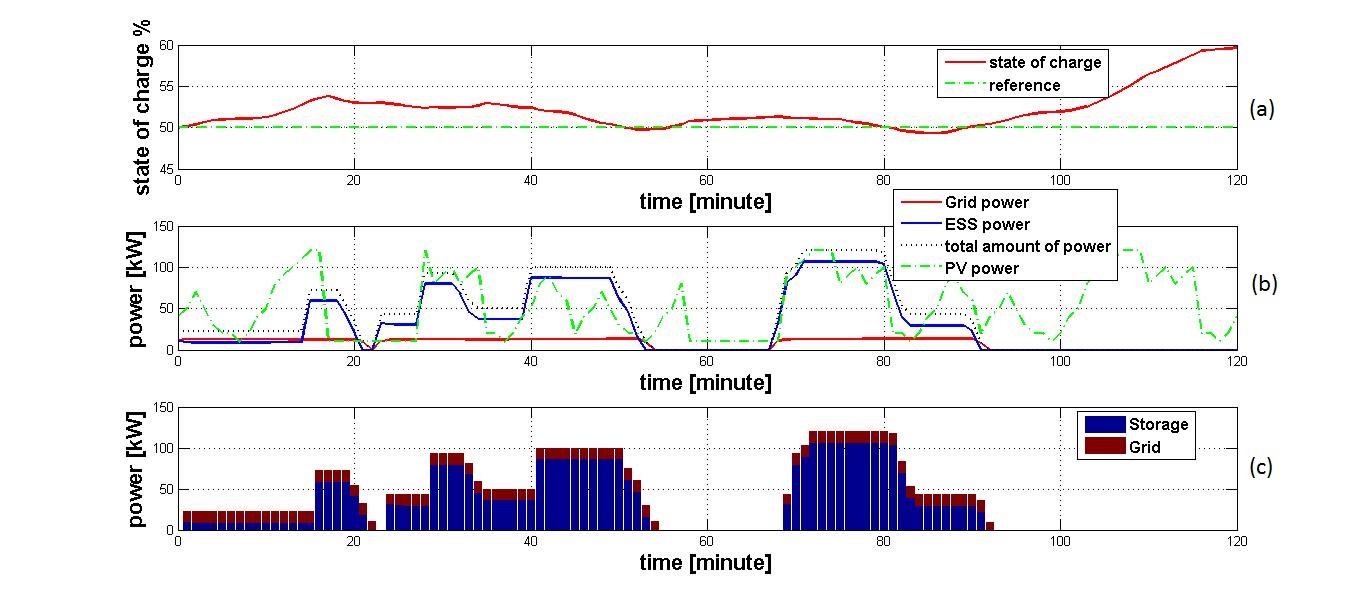}
	\vspace{-0.8cm}
	\caption{Grid connected system in time dependent case with medium level of storage charge and high level of $\delta$: \textit{(a)} ESS's state of charge and reference, \textit{(b)} Grid, ESS and PV power and total power, \textit{(c)} ESS and grid power}
	\label{fig:high_delta}
\end{figure}
\begin{figure}[h!]
	\centering
	\includegraphics[clip,width=1\columnwidth]{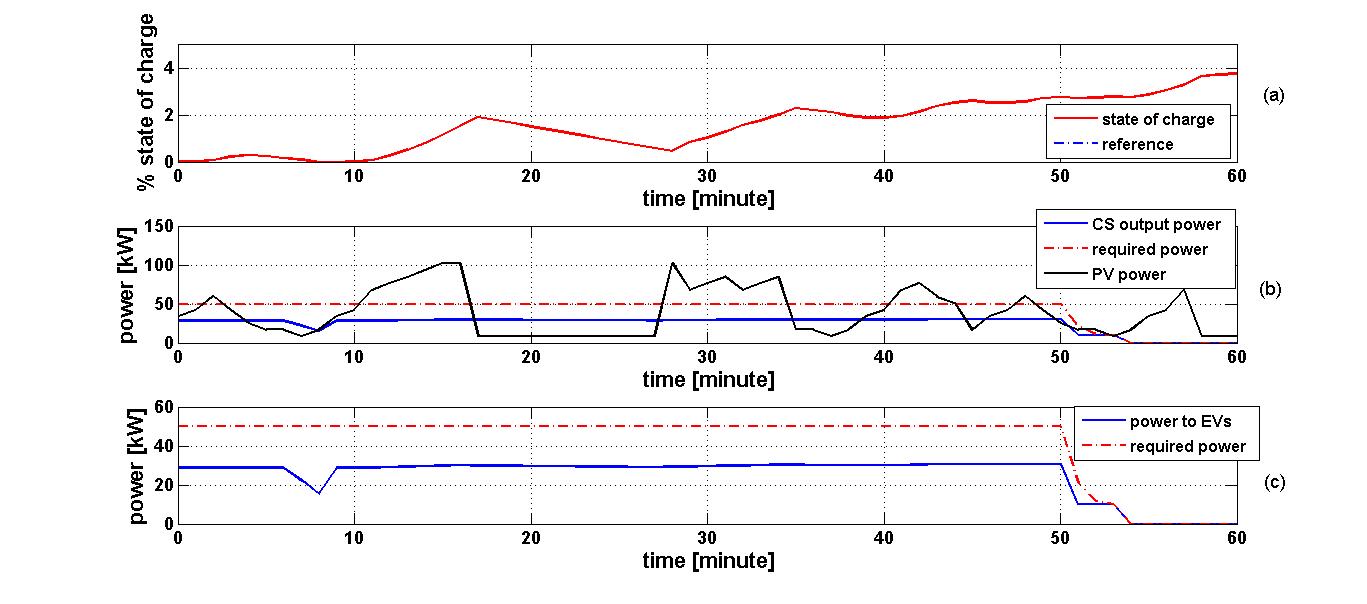}
	\vspace{-0.8cm}
	\caption{Stand alone system in time independent case with storage initially fully discharged: \textit{(a)} ESS's state of charge and reference, \textit{(b)} FCS, PV power and EV power reference, \textit{(c)} EV power reference and actual charging power.}
	\label{fig:sa_ind}
\end{figure}
\begin{figure}[h!]
\label{sa_dep}
	\centering
	\includegraphics[clip,width=0.99\columnwidth]{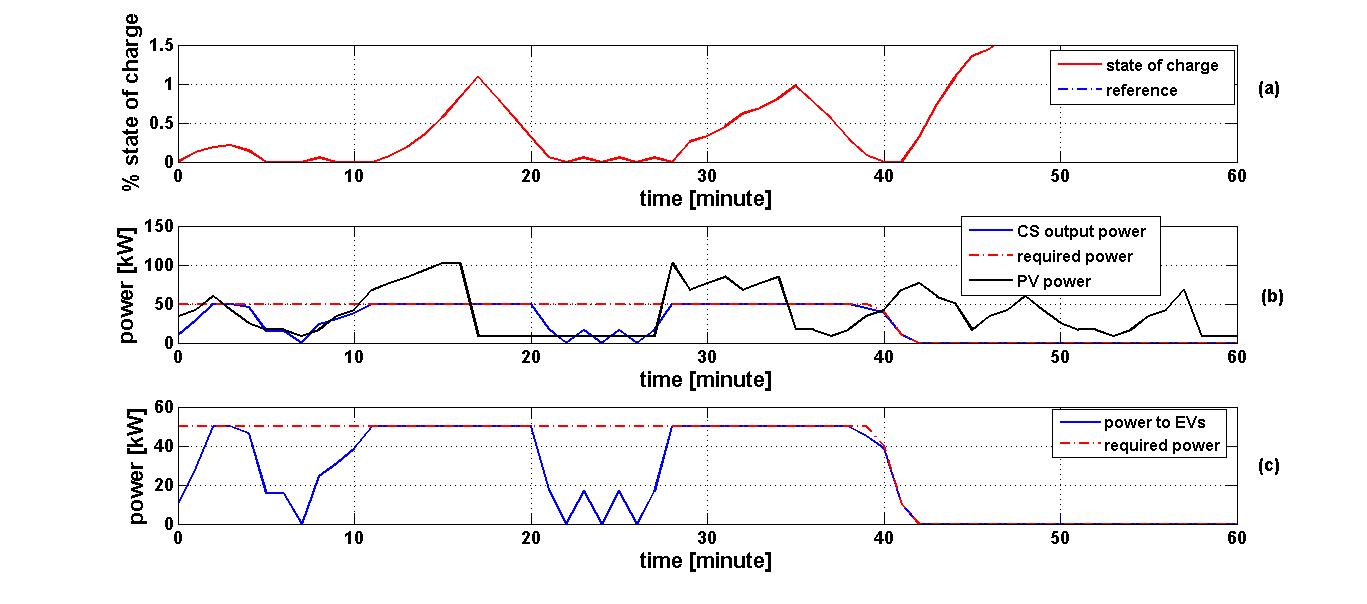}
	\vspace{-0.8cm}
	\caption{Stand alone system in time dependent case with storage initially fully discharged: \textit{(a)} ESS's state of charge and reference, \textit{(b)} FCS, PV power and EV power reference, \textit{(c)} EV power reference and actual charging power.}
	\label{fig:sa_dip}
\end{figure}
In the grid connected case the charging power profiles are equal to the references by design, so that the times of arrival have an impact on the generation of references only, as discussed in subsection \ref{sez:references}. Simulations of this use case have been focused on assessing the performance of the control strategy at POC level for different values of $\delta$ (fig.\ref{fig:low_delta} and fig.\ref{fig:high_delta}), taking into account the charging power references generated according to the time dependent strategy.  
In this set of simulations, it is not reported the power delivered to each vehicle but the aggregated one. The reference ESS's state of charge ($50 \%$) is considered again as initial value. 

Different values of $\delta$ produce the following implications: in those moments EVs charging operations require a significant change in the contribution from the grid, the power grid profile has a smooth shape in case $\delta$ is high, while has a steep profile in case $\delta$ is low, as it is possible to note for example at the minute 22 in both figures, as expected. At the same time, a high level of $\delta$ does not encourage the system to vary the contribution of the grid which remains constant as much as possible. For this reason, by comparing the two situations, $P^{g}$ is greater when $\delta$ is high.

\subsection{Effect of time dependent component on EV against ESS operation priority} \label{sez:caso_particolare}
In this subsection another implication of the presence of the temporal component is analyzed, taking as reference the stand alone plant. In particular, the following case is represented and depicted in fig.\ref{fig:sa_ind} and fig.\ref{fig:sa_dip}: an EV fully discharged arrives at the FCS and connects to the  50 kW plug when the ESS state of charge is equal to zero.
In the time independent case, the power provided by the PV is mainly used to recharge the storage, extending EV charging time over acceptable limits.
When the time component is considered, its presence contributes to shift the weight on the EVs charging term.
Since each EV needs a minimum power to be recharged ($\check{P}_m$), as soon as the ESS stores the minimum level of charge to release $\check{P}_m$, this minimum power level is fully used to recharge the EV. In this way, the risk of power fluctuations is high, as in the time period between minutes 22 and 25, and this intermittent behavior could cause component damage and malfunctions on the charging process. 
By comparing the two situations, it is possible to note that the time component does not affect only the strategy of charging in presence of several EVs: even when there is only one vehicle, the growth of the term $(t-t^{arr})^e$ during charging has an effect, forcing the system to encourage the charging of the vehicle at the expense of the storage.
\section{Conclusions and future works} 
\label{sez:conclusions}
In this paper a time driven MPC approach for the management
of EVs fast charging in a service area equipped with PV generation and a storage device has been presented. Two different use cases have been considered: in the former the charging area is disconnected from the grid, in the latter the grid contributes to EVs charging operations. A primary references generation problem has been analyzed in order to satisfy the constraint related to the maximum power delivered by the FCS, followed by a redefinition of set points depending on ESS's state of charge (only in stand alone case). Temporal dependancy has been introduced in the control strategy and its implications discussed, both in reference generation and charging operations tasks. Simulations have been performed for both use cases considering simultaneous charging sessions, which show the effectiveness of the proposed approach in integrating the ESS in fast charging applications. Also a special case placed in the stand alone plant has been discussed, in which a single EV has been considered in order to assess how the temporal component affects the priority in ESS and EVs charging operations. 

Future works consider the optimal sizing of each involved element and the integration of controllable RES in the control strategy. On one hand the sizing of grid connection represents a critical cost item, so that an optimal assignment of values to the parameters and a right sizing of storage capacity may lead to a substantial decreasing of grid contribution. On the other it is worth of investigation integrating the proposed EVs charging strategy with wind power control \cite{dg} or with residential energy aware networks \cite{VS1} \cite{VS2}. Also an interesting direction for research considers reliable integration of distributed FCSs in virtual areas, taking advantage of practical and theoretical results related to critical infrastructure protection \cite{FDP1}\cite{FDP2}\cite{vitos}\cite{AP2} and distributed control~\cite{AP1}\cite{AP3}.

%The above-mentioned problems are strictly related to the problem of critical infrastructure control, extensively dealt with in the literature ranging from theoretical studies [A], [B] to practical solutions [C], [D]. More specifically, optimal location and size of FCSs contribute to cost reductions....

%[...] it is worth of investigation integrating the proposed EVs charging strategy with wind power control [16] or with residential energy aware networks ([17] and [18]).

%%%%%%%%%%%%%%%%%%%%%%%%%%%%%%%%%%%%%%%%%%%%%%%%%%%%%%%%%%%%%%%%%%%
% ACKNOWLEDGMENT
%\section*{Acknowledgment}
%The authors would like to thank Prof. Francesco Delli Priscoli for the provisioning of relevant data and the helpful suggestions.

%%%%%%%%%%%%%%%%%%%%%%%%%%%%%%%%%%%%%%%%%%%%%%%%%%%%%%%%%%%%%%%%%%%%%%
%REFERENCES
%\vspace{0.48 cm}

\bibliographystyle{myIEEEtran}
\bibliography{science}

\end{document}